\documentclass[fleqn]{wlscirep}

\usepackage{xcolor}
\usepackage{blkarray}
\usepackage{listings}

\title{Enduring Lagrangian coherence of a Loop Current ring assessed using
independent observations}

\author[1,*]{Francisco J. Beron-Vera}
\author[2]{Mar\'{\i}a J. Olascoaga}
\author[3]{Yan Wang}
\author[4,**]{Joaqu\'{\i}n Tri\~nanes}
\author[5]{Paula P\'erez-Brunius}

\affil[1]{Department of Atmospheric Sciences, Rosenstiel School of
Marine and Atmospheric Science, University of Miami, Miami, Florida,
USA.} \affil[2]{Department of Ocean Sciences, Rosenstiel School of
Marine and Atmospheric Science, University of Miami, Miami, Florida,
USA.} \affil[3]{Department of Atmospheric and Ocean Sciences,
University of California Los Angeles, Los Angeles, California, USA.}
\affil[4]{Atlantic Oceanic and Atmospheric Laboratory, National
Oceanic and Atmospheric Administration, Miami, Florida, USA.}
\affil[5]{ Centro de Investigaci\'on Cient\'{\i}fica y de Educaci\'on
Superior de Ensenada, Ensenada, Baja California, M\'exico.}
\affil[*]{Corresponding author.  E-mail: fberon@rsmas.miami.edu.}
\affil[**]{Also at Cooperative Institute for Marine and Atmospheric
Studies, University of Miami,  Miami, Florida, USA; and Instituto
de Investigaciones Tecnol\'ogicas, Universidad de Santiago de
Compostela, Santiago, Espa\~na.}

\date{\today}


\begin{abstract}
  Ocean flows are routinely inferred from low-resolution satellite
  altimetry measurements of sea surface height assuming a geostrophic
  balance. Recent nonlinear dynamical systems techniques have
  revealed that surface currents derived from altimetry can support
  mesoscale eddies with material boundaries that do not filament
  for many months, thereby representing effective transport mechanisms.
  However, the long-range Lagrangian coherence assessed for mesoscale
  eddy boundaries detected from altimetry is constrained by the
  impossibility of current altimeters to resolve ageostrophic
  submesoscale motions.  These may act to prevent Lagrangian coherence
  from manifesting in the rigorous form described by the nonlinear
  dynamical systems theories. Here we use a combination of satellite
  ocean color and surface drifter trajectory data, rarely available
  simultaneously over an extended period of time, to provide
  observational evidence for the enduring Lagrangian coherence of
  a Loop Current ring detected from altimetry.  We also seek
  indications of this behavior in the flow produced by a data-assimilative
  system which demonstrated ability to reproduce observed relative
  dispersion statistics down into the marginally submesoscale range.
  However, the simulated flow, total surface and subsurface or
  subsampled emulating altimetry, is not found to support the
  long-lasting Lagrangian coherence that characterizes the observed
  ring.  This highlights the importance of the Lagrangian metrics
  produced by the nonlinear dynamical systems tools employed here
  in assessing model performance.
\end{abstract}

\begin{document}

\flushbottom
\maketitle

\thispagestyle{empty}

\section*{Introduction}

The prevalent looping behavior of satellite-tracked surface drifting
buoys and abundance of persistently closed streamlines of the
satellite altimetry sea surface height (SSH) field suggest that
eddies with water trapping capability are ubiquitous in the ocean
\cite{Chelton-etal-11a, Lumpkin-15}.  However, owing to the
observer-dependent nature of position and velocity, this is only a
loose and generally incorrect assessment of Lagrangian (i.e.,
material) coherence \cite{Haller-05, Beron-etal-13, Serra-Haller-16,
Rempel-etal-17}.

To assess Lagrangian coherence irrespective of the observer, a
specialized technique is necessary. Recent developments in the area
of nonlinear dynamical systems have led to such a technique, enabling
objective detection from a finite-time-aperiodic flow realization
of eddies with material boundaries that experience no filamentation
over the detection interval \cite{Haller-Beron-12, Beron-etal-13,
Haller-Beron-13, Haller-Beron-14, Haller-etal-16}.  Detecting eddies
with that property is important because of the impact they may have
in global ocean transport \cite{Lehahn-etal-11, Wang-etal-15,
Wang-etal-16, Abernathey-Haller-18}.

Long-lived \emph{coherent Lagrangian eddies} have been extracted
from flows derived geostrophically from satellite altimetric
measurements of SSH in a number of occasions \cite{Beron-etal-15,
Romero-etal-16, Wang-etal-15, Wang-etal-16}.  While satellite
altimetry is widely used to monitor global ocean variability
\cite{Beron-etal-08b, Beron-etal-10b, Fu-etal-10}, the long-range
Lagrangian coherence assessed for mesoscale eddies detected from
altimetry is constrained by the impossibility of current altimeters
to resolve submesoscale processes, which are increasingly believed
to be important in the upper ocean \cite{McWilliams-08a, Klein-09,
Poje-etal-14}.  While the mesoscale dynamics still play a dominant
stirring role \cite{Olascoaga-etal-13, Beron-LaCasce-16, Haza-etal-16,
Gough-etal-17, Duran-etal-18}, submesoscale processes may contribute
to prevent Lagrangian coherence from manifesting in the rigorous
form predicted by the nonlinear dynamical systems theories.

The goal of this paper is to provide support for the enduring
Lagrangian coherence of a mesoscale eddy using an unprecedented
combination of observations from independent sources.  We first
detect from altimetry an exceptionally persistent coherent Lagrangian
Loop Current ring (LCR) in the Gulf of Mexico (GoM).  Then we show
that associated with the ring is a satellite-derived chlorophyll
deficient patch, confirming the material nature of the ring.  Using
satellite-tracked surface drifter trajectories we provide evidence
that the boundary of the ring is largely convex (wiggle free),
showing no significant signs of the presence of submesoscale eddies
rolling up around its material boundary.

We also analyze the output from the operational U.S. Navy Coastal
Ocean Model (NCOM) of the GoM.  At 1-km-horizontal resolution, this
data-assimilative system marginally resolves submesoscales.  We
show that the neither the velocity field derived by subsampling the
model's SSH simulating satellite altimetry nor the total upper-ocean
model velocities supports a coherent Lagrangian ring with similar
characteristics.  The simulated altimetry flow produces a smaller
and much shorter lived coherent Lagrangian ring, and the total model
flow does not sustain a coherent Lagrangian ring, neither at the
surface nor below of it within the first 50~m.  The latter, however,
do reveal the presence of swirling Lagrangian structures that are
coherent but in a relaxed sense recently proposed
\cite{Beron-etal-18a}. [The work of Beron-Vera et al.\
\cite{Beron-etal-18a} relies to a large extent on the results from
the present paper, which was submitted for publication elsewhere
several months before Beron-Vera et al.'s \cite{Beron-etal-18a} work
was submitted. Beron-Vera et al. \cite{Beron-etal-18a} reference an
earlier version of this paper under title ``On the significance of
coherent Lagrangian eddies detected from satellite altimetry''
posted by the authors in arXiv:170406186.]
   
\section*{Methods}

\subsection*{Coherent Lagrangian eddies}

\begin{figure}[p!]
  \centering%
  \includegraphics[width=\textwidth]{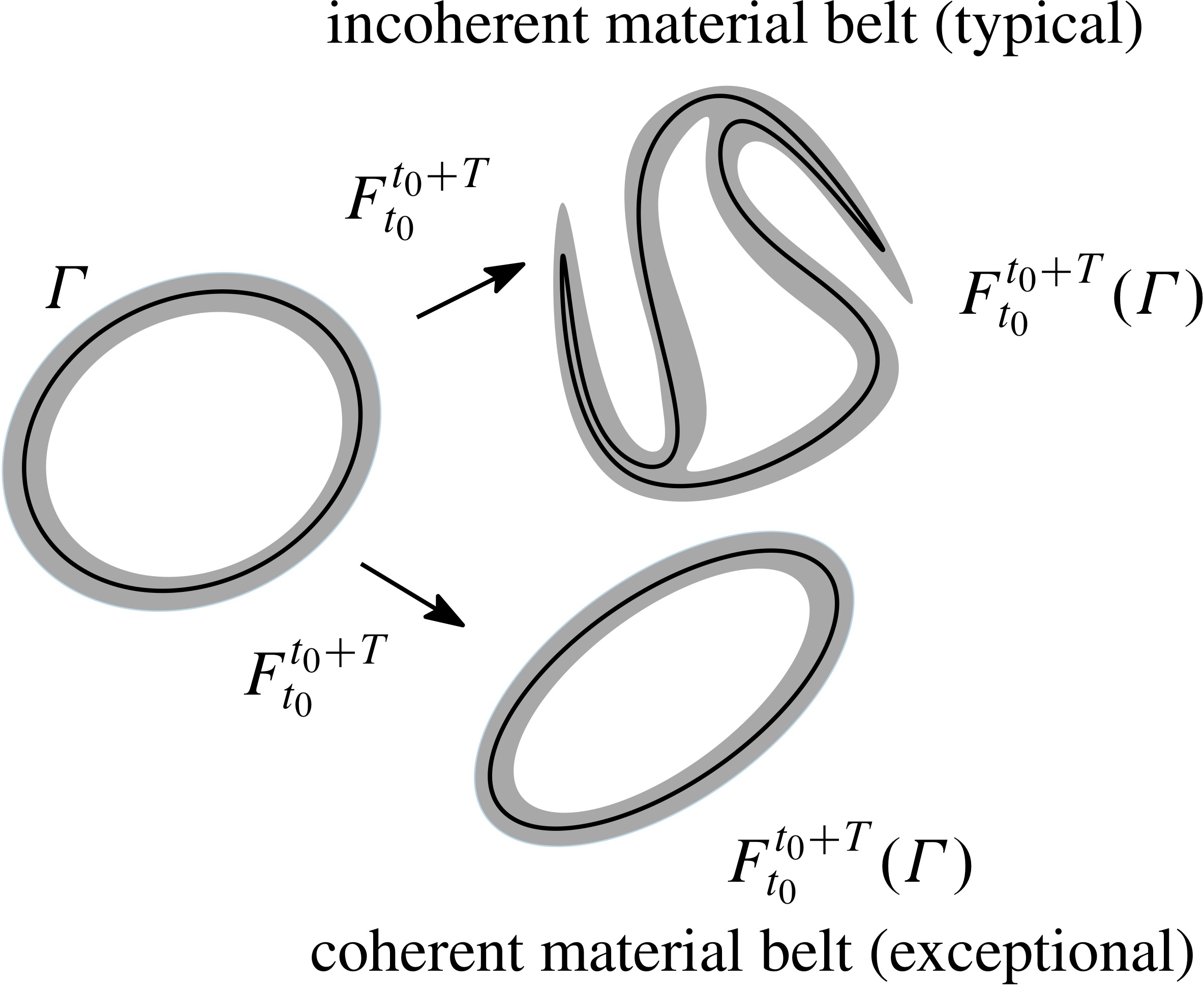}%
  \caption{A material loop $\Gamma$ (black) at time $t = t_0$ is
  advected by a two-dimensional flow into its later position
  $\smash{F^{t+T}_{t_0}(\Gamma)}$ at time $t = t_0+T$. The advected
  material loop remains coherent if an initially thin material belt
  (grey) around the material loop experiences no leading-order
  variations in averaged material stretching upon advection. The
  outermost member in a family of nonintersecting such material
  loops constitutes the optimal boundary of a coherent material
  eddy.  Figure constructed using Ipe (http://ipe\allowbreak
  .otfried.\allowbreak org/).}
  \label{fig:belts}%
\end{figure}

Haller and Beron-Vera \cite{Haller-Beron-13, Haller-Beron-14} seek
fluid regions enclosed by exceptional material loops that defy the
exponential stretching that a typical loop will experience in
turbulent flow.  As demonstrated by these authors, such loops, which
constitute a type (elliptic) of Lagrangian coherent structures
\cite{Haller-15}, have small annular neighborhoods exhibiting no
leading-order variation in averaged material stretching (Fig.\
\ref{fig:belts}).

Solving the above variational problem reveals that the loops in
question are uniformly stretching: any of their subsets are stretched
by the same factor $\lambda$ under advection by the flow from time
$t_0$ to time $t$.  The time $t_0$ positions of $\lambda$-stretching
material loops turn out to be limit cycles of one of the following
two equations for parametric curves $s \mapsto x_0(s)$:
\begin{equation}
  \frac{\mathrm{d}x_0}{\mathrm{d}s} = 
  \sqrt{
  \frac
  {\lambda_2(x_0) - \lambda^2}
  {\lambda_2(x_0) - \lambda_1(x_0)}
  }
  \,\xi_1(x_0) 
  \pm
  \sqrt{
  \frac
  {\lambda^2 - \lambda_1(x_0)}
  {\lambda_2(x_0) - \lambda_1(x_0)}
  }
  \,\xi_2(x_0),  
  \label{eq:eta}
\end{equation}
where $\lambda_1(x_0) < \lambda^2 < \lambda_2(x_0)$.  Here
$\{\lambda_i(x_0)\}$ and $\{\xi_i(x_0)\}$, satisfying $0 <
\lambda_1(x_0) \equiv \smash{\lambda_2(x_0)^{-1}} < 1$, $\xi_i(x_0)
\cdot \xi_j(x_0) = \delta_{ij}$ ($i,j = 1,2$), are eigenvalues and
(normalized) eigenvectors, respectively, of the Cauchy--Green tensor,
$\smash{C_{t_0}^t(x_0)} := \smash{\mathrm{D}F_{t_0}^t(x_0)^\top
\mathrm{D}F_{t_0}^t(x_0)}$, an objective (i.e., observer-independent)
descriptor of material deformation, where $\smash{F_{t_0}^t} : x_0
\mapsto x(t;x_0,t_0)$ is the flow map that takes time-$t_0$ positions
on a horizontal plane to time-$t$ positions of fluid particles,
which obey
\begin{equation}
  \frac{\mathrm{d}x}{\mathrm{d}t} = v(x,t),
  \label{eq:dxdt}
\end{equation}
where $v(x,t)$ is a two-dimensional velocity field.

Limit cycles of (\ref{eq:eta}) either grow or shrink under changes
in $\lambda$, forming smooth annular regions of nonintersecting
loops. The outermost member of such a band of material loops is
observed physically as the boundary of a \emph{coherent Lagrangian
eddy}.  Limit cycles of (\ref{eq:eta}) tend to exist only for
$\lambda \approx 1$.  Material loops characterized by $\lambda =
1$ resist the universally observed material stretching in turbulence:
they reassume their initial arclength at time $t$.  This conservation
of arclength, along with enclosed area preservation, can produce
extraordinary coherence \cite{Beron-etal-13}.  Finally, limit cycles
of (\ref{eq:eta}) are (null) geodesics of the generalized Green--Lagrange
tensor $\smash{C_{t_0}^t}(x_0) - \lambda^2\mathop{\rm Id}$, which must
necessarily contain degenerate points of $\smash{C_{t_0}^t}(x_0)$
where its eigenvector field is isotropic.  For this reason the above
procedure is known as \emph{geodesic eddy detection}.

The numerical implementation of geodesic eddy detection is documented
at length \cite{Haller-Beron-12, Haller-Beron-13, Hadjighasem-etal-13,
Haller-Beron-14, Karrasch-etal-14, Beron-etal-15, Haller-15,
Wang-etal-15, Hadjighasem-Haller-16, Wang-etal-16}.  A software
tool is also available \cite{Onu-etal-15}.  Here we have set the
grid spacing of the computational domain to 0.1 km \cite{Olascoaga-etal-13,
Wang-etal-15, Wang-etal-16, Duran-etal-18, Beron-etal-18a}.  All
integrations were carried out using a step-adapting fourth/fifth-order
Runge--Kutta method with interpolations done with a cubic
method.

We close this section by noting that while geodesic eddy detection
is two-dimensional in nature, it can be applied (as done below)
levelwise two-dimensionally to gain insight into three-dimensional
aspects of the flow as described for instance by a hydrostatic model
like NCOM.

\subsection*{Data} 

The SSH is taken as the sum of a (steady) mean dynamic topography
and the (transient) altimetric SSH anomaly.  The mean dynamic
topography is constructed from satellite altimetry data, in-situ
measurements, and a geoid model \cite{Rio-Hernandez-04}.  The SSH
anomaly is provided weekly on a 0.25$^{\circ}$-resolution
longitude--latitude grid.  This is referenced to a 20-year (1993--2012)
mean, obtained from the combined processing of data collected by
altimeters on the constellation of available satellites
\cite{LeTraon-etal-98}.

The ocean color data consist of chlorophyll concentration fields
obtained from the National Aeronautics and Space Administration
(NASA) Biology Processing Group website (http://oceancolor.\allowbreak
gsfc.\allowbreak nasa.\allowbreak gov).  Sun glint and clouds
represent a major source of data gaps in the Gulf of Mexico,
especially during the summer months. To minimize their effects and
improve coverage,  we have chosen to use multiday Level-3 mapped
products.  These are 8-day composite chlorophyll fields at 4-km
horizontal resolution from the MODIS (Moderate Resolution Imaging
Spectroradiometer) sensor, on-board NASA's Earth Observing System
(EOS)  \emph{Terra} and \emph{Aqua} satellites, and the VIIRS
(Visible Infrared Imaging Radiometer Suite) sensor, aboard the
\emph{Suomi-NPP} satellite.

The surface drifters are Far Horizon Drifters (FHD), manufactured
by Horizon Marine, Inc. \cite{Sharma-etal-10, Anderson-Sharma-08}.
These are deployed by Horizon Marine Inc. as part of the
EddyWatch\textsuperscript{\textregistered} program .  A subset of
7 to 16 drifters deployed within or in the closest vicinity of the
LCR under investigation are considered. Each drifter consists of a
cylindrical buoy attached to a 45-m tether line, attached itself
to a 1.2-m ``para-drogue.'' These instruments are deployed by air,
so the drogue serves both as parachute to protect the buoy when air
deployed, and to reduce wind slippage of the buoy as it drifts in
the water.  Positions are recorded hourly using GPS that transmit
the data via the \emph{Argos} system. The quality control of the
data consisted of interpolating positions between data gaps shorter
than 6 hours, and removing locations over land and data points with
speeds exceeding 3 m s$^{-1}$.  Correlations of the drifter velocities
with winds \cite{Brunius-etal-13} are of the same order as those
obtained with the SVP drifters \cite{Poulain-etal-09}, which are
drogued with holey socks at about 15 m. This suggests that wind-slippage
on the FHD drifter is minimal and the water-following characteristics
of FHD drifters are in general similar to the drifters used by the
National Oceanic and Atmospheric Administration (NOAA) Global Drifter
Program \cite{Lumpkin-Pazos-07}.

\subsection*{Model} 

The NCOM simulation employs assimilation and nowcast analyses from
NCODA (Navy Coupled Ocean Data Assimilation) \cite{Cummings-05}.
Forecasts are generated by systems linking NCODA with regional
implementations \cite{Rowley-Mask-14} of NCOM \cite{Barron-etal-06}.
The model has 1-km horizontal resolution and was initiated on 15
May 2012 from the then operational global ocean model Global Ocean
Forecast System (GOFS) 2.6 \cite{Barron-etal-07}.  Daily boundary
conditions are received from the current operational GOFS using the
HYbrid Coordinate Ocean Model (HYCOM) \cite{Metzger-etal-09}.  The
vertical grid is comprised of 49 total levels; 34 terrain-following
$\sigma$-levels above 550 m and 15 lower $z$-levels. The
$\sigma$-coordinate structure has higher resolution near the surface
with the surface layer having 0.5-m thickness. The simulation uses
atmospheric forcing at the sea surface from COAMPS (Coupled
Ocean/Atmosphere Mesoscale Prediction System) \cite{NRL-97} to
generate forecasts of ocean state up to 72 h in 3-hour increments.
The observational data assimilated in these studies is provided by
NAVOCEANO (Naval Oceanographic Office) and introduced into NCODA
via its ocean data quality control process.  Observations are
three-dimensional variational (3DVar) assimilated \cite{Smith-etal-11}
in a 24-hour update cycle with the first guess from the prior day
NCOM forecast.

\section*{Results}

The focus of our analysis is a region filled with closed streamlines
of the altimetric SSH field in the GoM, which can be tracked for
nearly one year since April 2013.  The region was classified as an
LCR by the Horizon Marine Inc.'s EddyWatch\textsuperscript{\textregistered}
program (http://\allowbreak www.\allowbreak horizonmarine.\allowbreak
com) using Leben's \cite{Leben-05} methodology, which searches for
Eulerian footprints of LCRs on the SSH field, and further named
\emph{Kraken}.

\begin{figure}[p!]
  \centering%
  \includegraphics[width=\textwidth]{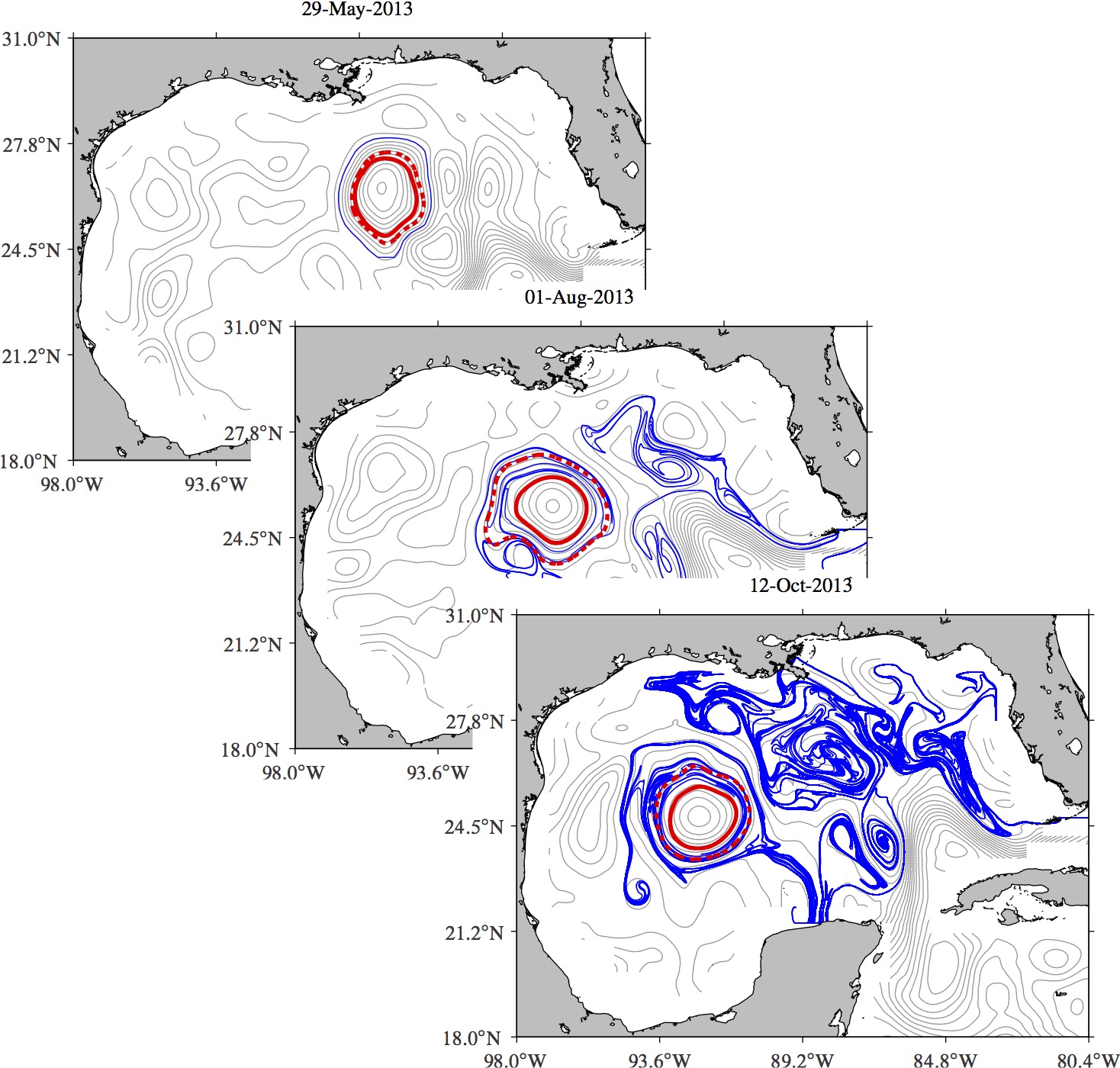}%
  \caption{In grey, contour levels of the altimetric sea surface
  height (SSH) field in the Gulf of Mexico (GoM) on selected days.
  Overlaid in blue are snapshots of the passive evolution, according
  to the altimetry-derived flow, of tracers initially on 29 May
  2013 along the outermost of the closed SHH streamlines filling a
  mesoscale region of nearly 100-km radius in the center of the
  GoM, which has been identified as a Loop Current ring (LCR) and
  named \emph{Kraken}. These closed instantaneous SSH streamlines
  are the Eulerian footprints of a coherent Lagrangian LCR, which
  nearly steadily translates westward across the GoM.  The solid
  red curve is the boundary for the longest-lived and largest
  coherent Lagrangian LCR core. The dashed red curve is a material
  loop that provides repeated, short-term shielding to this core.
  Figures constructed using MATLAB R2017b (http://www\allowbreak
  .mathworks.\allowbreak com/).}
  \label{fig:ssh}%
\end{figure}

Figure \ref{fig:ssh} shows in grey several contour levels of the
SSH field on three selected days, progressing from top to bottom.
Highlighted in blue on the top is the outermost closed SSH contour
in the region identified as LCR \emph{Kraken} on the indicated day.
Two images of this curve under the flow map associated with the
altimetry-derived advection field are shown on the middle and bottom.
More specifically, these are the result of integrating the passive
tracer particle equation (\ref{eq:dxdt}), with the velocity derived
geostrophically from SSH, from tracer positions initially on 29 May
2013 along the curve.  Several additional frames, including
instantaneous SSH isolines as well as advected images of the tracer
curve over several months, are shown in Supplementary Movie S1.

Note the filamentation experienced by the passively advected tracer
curve.  Note also that, while intense, the filamentation is mainly
outward.  The outward-filamented curve portion reveals a number of
swirling flow regions of varied sizes in the mesoscale range resolved
by altimetry and also below, as a result of nonlocal straining
by the altimetry-derived flow.  Most of these swirls are not seen
to preserve their entity over time.  The inward-filamented curve
piece, by contrast, reveals a enduring compact region, of a mean
radius of nearly 100 km, contained inside the region where the SSH
streamlines are instantaneously closed and has been been identified
as LCR \emph{Kraken}.  Two key observations about the behavior of
the tracer near this region are in order.  First, the tracer does
not invade the region over the course of many months.  Rather, it
wraps around the region in an anticyclonic fashion.  Second, the
wrapping tracer tends to occupy a thick belt (with a width of 10
to 20 km) around the region in question.  These two observations
suggest on one hand that LCR \emph{Kraken} has an impermeable
material core, surrounded by a resilient material loop which acts
as a long-term barrier for the transport of tracer into and out
from the material core.  On the other hand, outside this long-term
closed transport barrier, shorter-term material barriers must be
emerging repeatedly, shaping and confining the tracer wrapping
around the long-term coherent Lagrangian ring core.

Geodesic eddy detection has been designed to unveil the precise
location of such long- and short-term resilient material loops.
These are indicated by a solid and dashed red curve, respectively,
in Fig.\ \ref{fig:ssh}.  To unveil them, we applied geodesic eddy
detection on the altimetry-derived velocity field over the time
interval $[t_0, t_0+T]$ for appropriate choices of $t_0$ and $T$.

The specific choices $t_0 =$ 29 May 2013 and $T =$ 200 d were
found to produce a resilient material loop that provides the boundary
for the largest and longest-lived coherent Lagrangian ring core.
From $t_0 =$ 29 May 2013 to $t_0 + T =$ 15 December 2013, this
boundary uniformly stretches by a factor $\lambda = 1.05$.  This
minor uniform stretching of the boundary along with the conservation
of the enclosed area, which holds because the velocity field is
geostrophic and hence divergenceless, conveys LCR \emph{Kraken}
exceptional Lagrangian coherence.

Shorter-term resilient material loops encompassing the largest area
which includes the long-term coherent Lagrangian ring core are found
from the application of geodesic eddy detection every $t_0$ from
29 May daily through 15 December 2013 using $T =$ 30 d.  These
emerging material loops reassume their initial arclength (i.e.,
they have $\lambda = 1$) after a period of one month.  As such,
they convey additional temporary but repeated shielding to the ring
core, reinforcing its coherent Lagrangian nature.  Successive
short-term coherence regain events where first reported by Wang et
al.\ \cite{Wang-etal-16} for Agulhas rings in the South Atlantic.

Beyond the fact that our inference on the exceptional Lagrangian
coherence of LCR \emph{Kraken} remains valid irrespective of the
reference frame adopted by the observer, a unique aspect of geodesic
eddy detection, the pertinent question that arises is whether this
is a real ingredient of the ocean circulation in the GoM or is just
an artifact of the limitation of satellite altimetry to resolve
submesoscale and likely ageostrophic motions.  It turns out that,
as we proceed to show, LCR \emph{Kraken} constitutes to a large
extent a realistic coherent Lagrangian ring.  We support our
assertion with two pieces of independent observations, rarely
available simultaneously over an extended period of time.

\begin{figure}[p!]
  \centering%
  \includegraphics[width=\textwidth]{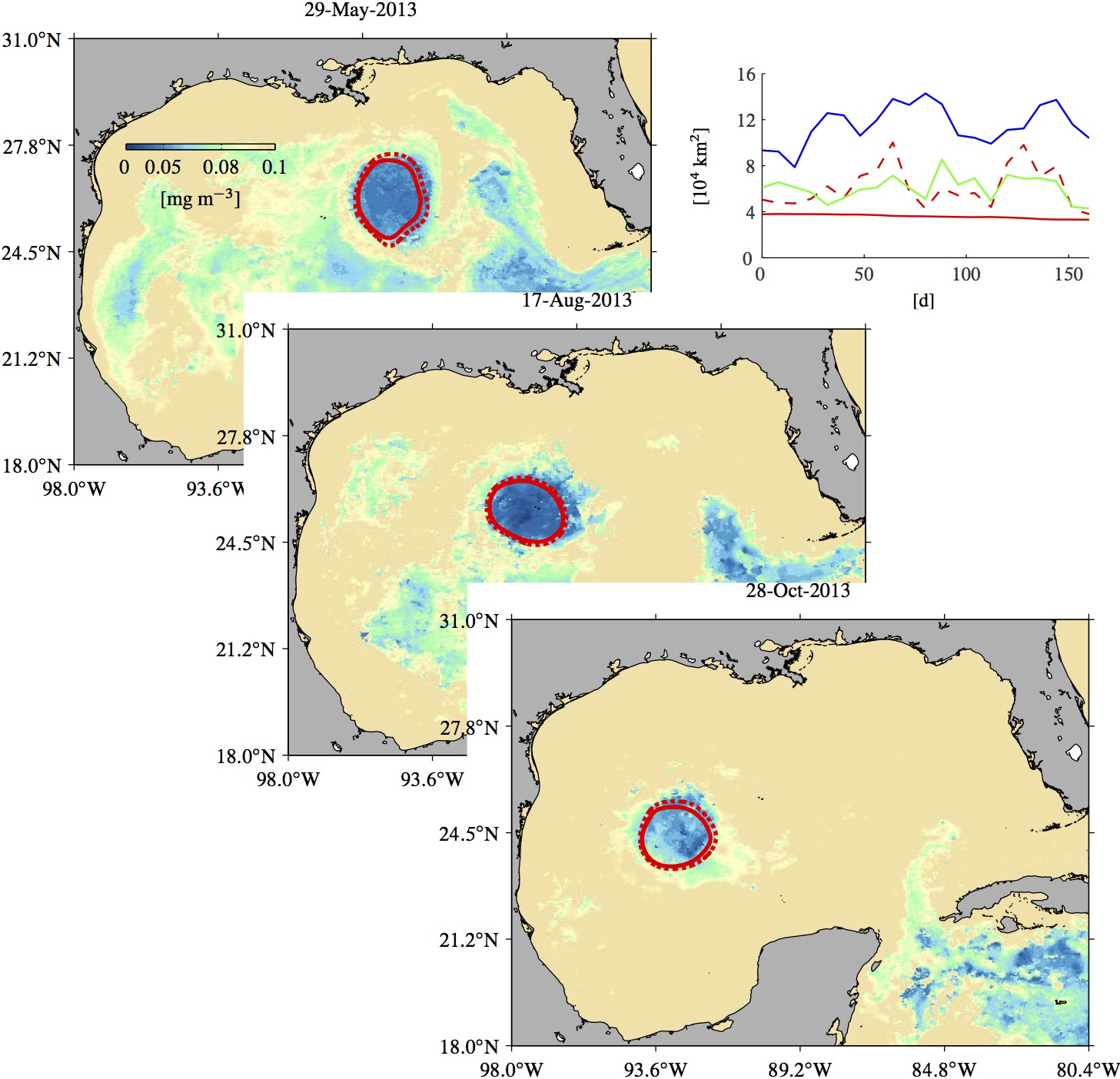}%
  \caption{Satellite-derived chlorophyll concentration images on
  selected days with the altimetry-inferred long- (solid) and
  short-term (dashed) coherent Lagrangian boundaries of LCR
  \emph{Kraken} overlaid. The top-right panel shows, as a function
  of time since 29 May 2013, area enclosed by the long- (solid red)
  and short-term (dashed red) Lagrangian boundaries, the outermost
  instantaneous closed SSH contour level (blue), and the area spanned
  by the chlorophyll deficient patch (green).  Figures constructed
  using MATLAB R2017b (http://www\allowbreak .mathworks.\allowbreak
  com/).}
  \label{fig:chl}%
\end{figure}

The first piece of observations is given by satellite ocean color.
Figure \ref{fig:chl} shows snapshots of chlorophyll concentration
on selected days with the altimetry-inferred long- and short-term
coherent Lagrangian ring boundaries overlaid (solid and dashed red,
respectively).  Weekly snapshots in the period 19 April through 21
November 2013 are shown in Supplementary Movie S2.

Note the patch of low chlorophyll concentration which, for a period
as long as about seven months, quite closely accompanies the
altimetry-inferred coherent Lagrangian ring as it nearly steadily
translates westward in the GoM.  Note also that the origin of the
chlorophyll defect can be traced completely into the Caribbean Sea,
which is characterized by predominantly oligotrophic conditions
\cite{Gomez-14}.  This indicates that the formation of LCR \emph{Kraken}
follows the standard scenario: the Loop Current occludes and LCR
\emph{Kraken} is subsequently pinched off from the current.  This
process is Lagrangian in essence \cite{Andrade-etal-13}.  Altogether
the two observations just made provide strong support to our
conclusion that LCR \emph{Kraken} is largely Lagrangian.

Clearly, some deviations between the boundaries of the low-chlorophyll
patch and the altimetry-inferred coherent Lagrangian ring are
evident.  But these are expected because of at least two reasons.
First, cloud coverage is important at times during the observational
period.  This results in nonuniform sampling, which impacts the
interpolation of the data on a regular grid and hence the resolution
of the chlorophyll defect.  Second, chlorophyll as a tracer cannot
be considered to be entirely passive.  Vertical pumping of nutrients,
not represented by the altimetry-derived flow, at the periphery of LCR
\emph{Kraken} may be expected \cite{Mahadevan-etal-08}.  Biological
activity stimulated by nutrient availability may be modifying the
chlorophyll concentration and thus the size of the patch as it
translates across the GoM.

However, the geodesically detected long- and short-term Lagrangian
boundaries much better constrain the extent of the chlorophyll
deficient patch than instantaneous closed streamlines of the SSH
field.  This is demonstrated in the top-right panel of Fig.\
\ref{fig:chl}, which compares, as a function of time since 29 May
2013, the area enclosed by the long- (solid red) and short-term
(dashed red) Lagrangian boundaries and the outermost instantaneous
closed SSH contour level (blue) with the area spanned by the
chlorophyll deficient patch (green), whose boundary was determined
as the chlorophyll isoline where the concentration change with
respect to the area enclosed by it is highest.  Note that the area
of the chlorophyll deficient patch is not constant as is (to numerical
error) that enclosed by the long-term boundary.  While biological
activity is likely acting to shape the form and extent of the patch,
this action is constrained by the Lagrangian circulation, which
favors coherence regain events that prevent the chlorophyll depleted
water from spreading away from LCR \emph{Kraken}.

\begin{figure}[p!]
  \centering%
  \includegraphics[width=\textwidth]{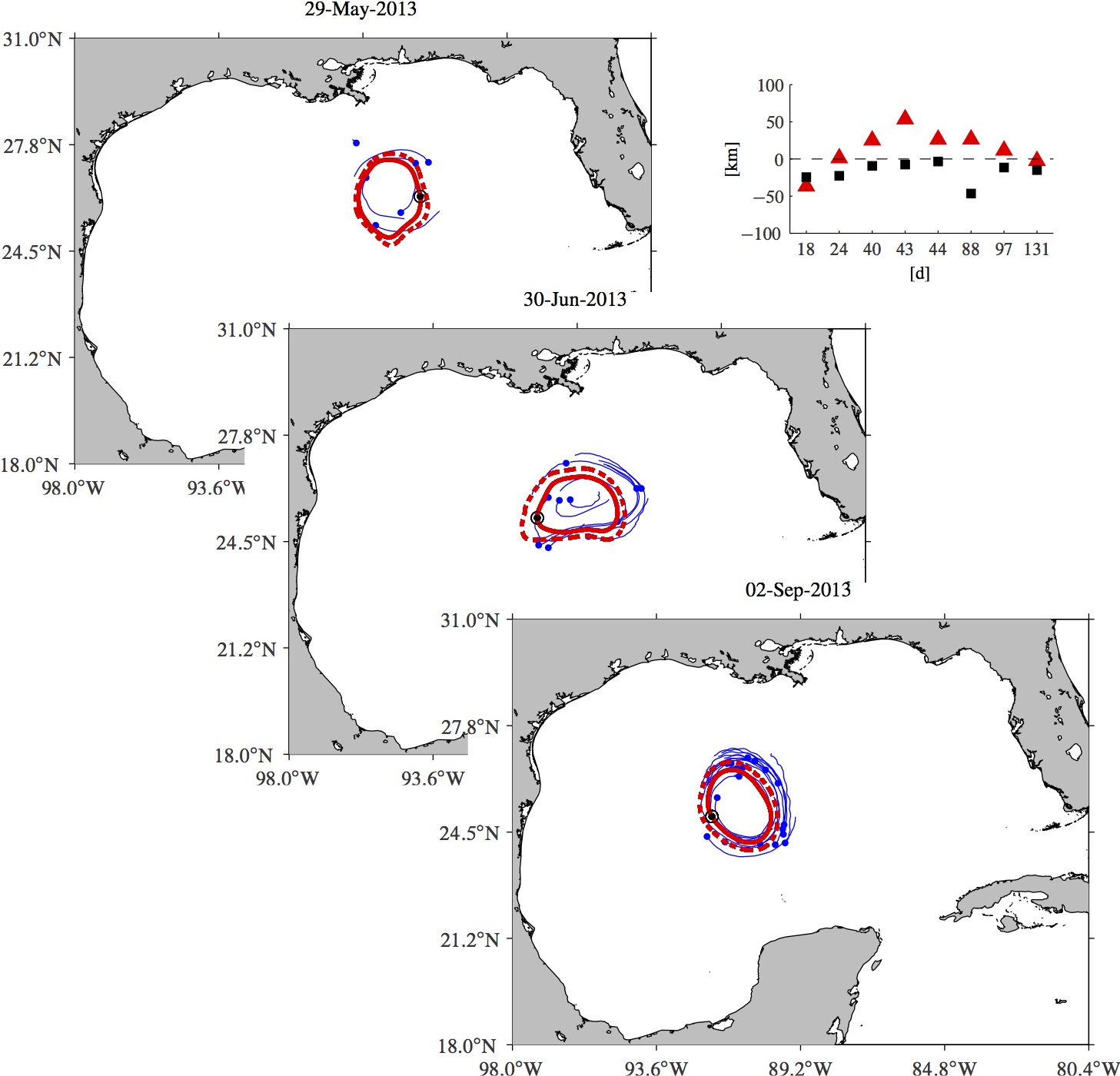}%
  \caption{Snapshots of the evolution of the altimetry-inferred
  coherent Lagrangian long- (solid red) and short-term (dashed red)
  boundaries of LCR \emph{Kraken} along with that of satellite-tracked
  surface drifters initially as close as 50 km from the long-term
  boundary (blue).  The drifter positions on each day shown are
  indicated by blue dots. The tails attached to the dots are week-long
  past trajectory segments.  One point on the long-term boundary
  is highlighted by a black dot surrounded by a circle, for reference.
  The top-right panel shows as a function of drifter lifespan,
  initial (black square) and mean (red triangle) shortest signed
  distance to the long-term boundary for all drifters initially
  lying within.   Figures constructed using MATLAB R2017b
  (http://www\allowbreak .mathworks.\allowbreak com/).}
  \label{fig:drifters}%
\end{figure}

The second piece of observations is provided by trajectories of
satellite-tracked surface drifters.  As part of the routine monitoring
of LCRs, the Horizon Marine Inc.'s
EddyWatch\textsuperscript{\textregistered} program deployed various
drifters in the vicinity of LCR \emph{Kraken} along its westward
path.  Figure \ref{fig:drifters} shows snapshots of the evolution
of the altimetry-inferred coherent Lagrangian ring long- and
short-term boundaries, indicated by solid and dashed red curves,
respectively, along with that of drifters initially lying within
these boundaries or which at some point in time have come as close
as 50 km to the long-term boundary (various additional snapshots
are shown in Supplementary Movie S3).  One point on the long-term
boundary is highlighted by a black dot surrounded by a circle, for
reference.  The drifter positions are indicated by blue dots on the
specified day.  The tails (in blue) attached to the dots are week-long
past trajectory segments.

The first aspect to note is that the drifters deployed inside or
near by the long- and short-term boundaries tend to remain in their
vicinity, very closely following the anticyclonic circulatory motion
of passive tracers along them.  This can be most easily visualized
by comparing the position of the point highlighted on the long-term
boundary and those of the nearby drifters.  The turnover timescale
of that point is of about two weeks, which is nearly the same for
the drifters in its vicinity.  Overall the behavior of the drifters
is largely consistent with the motion of tracer particles passively
advected by the altimetry-derived flow.  This provides further
support to the Lagrangian nature of LCR \emph{Kraken} inferred from
the analysis of altimetry data.

A quantitative assessment of the above qualitative observations is
provided in the top-right panel of Fig.\ \ref{fig:drifters}, which
shows, as a function of drifter lifespan, initial (black square)
and mean (red triangle) shortest signed distance to the long-term
coherent Lagrangian boundary of LCR \emph{Kraken} for all drifters
initially lying within (a negative value indicates that the drifter
is inside the boundary).  Two out of 8 drifters spent their whole
lifespan (18 and 131 d) within the long-term boundary. The other 6
drifters remained around the ring with a mean minimum distance
shorter than 50 km.

Inertial effects (i.e., of objects' finite size and buoyancy) might
explain the behavior of the relatively slow outward moving drifters.
Indeed, floating objects are predicted to be repelled away from
anticyclonic coherent Lagrangian eddies \cite{Beron-etal-15,
Haller-etal-16}.  These effects can be expected to be most effective
when the para-drogues attached to the drifters are closed.  The
outward motion of the drifters may be indicative of this.  Windage
may also explain deviations of drifter motion from purely Lagrangian
motion.  However, windage effects near mesoscale eddies or rings
have not been well constrained for biological tracers
\cite{McGillicuddy-etal-98, McGillicuddy-etal-07, Mahadevan-etal-08},
and they may be outbalanced by inertial effects \cite{Beron-etal-16}.

An additional important aspect to be highlighted of the behavior
of the drifters is the relative absence of submesoscale wiggles in
their hourly-sampled trajectories.  The lack of such wiggles in the
drifter trajectories is largely consistent with the convexity of
the altimetry-inferred long- and short-term coherent Lagrangian
ring boundaries.  This observation gives further support to the
significance of the altimetry-based inference.
 
It might be argued that the dominance of wiggle-free drifter
trajectories is a consequence of the drifters being para-drogued
at about 50-m depth, which lies somewhat below the summertime
mixed-layer base \cite{Muller-etal-15}.  However, this should not
prevent the drifters from sampling submesoscale motions if they
were active \cite{Brunius-etal-13, Lumpkin-15}.  Indeed, loops of
about 10-km or smaller diameter can be seen at times mounted on the
drifter trajectories.  But the frequency of those loops is typically
very close to the local Coriolis frequency, as it has been observed
earlier in the same region using the same drifters
\cite{Anderson-Sharma-08}.  So the loops mainly correspond to
near-inertial oscillations, which do not impact relative dispersion
statistics \cite{Beron-LaCasce-16} and thus should not be expected
to constitute an effective mechanism for the erosion of mesoscale
coherent Lagrangian structures such as the resilient material
boundary around LCR \emph{Kraken}.

\begin{figure}[p!]
  \centering%
  \includegraphics[width=\textwidth]{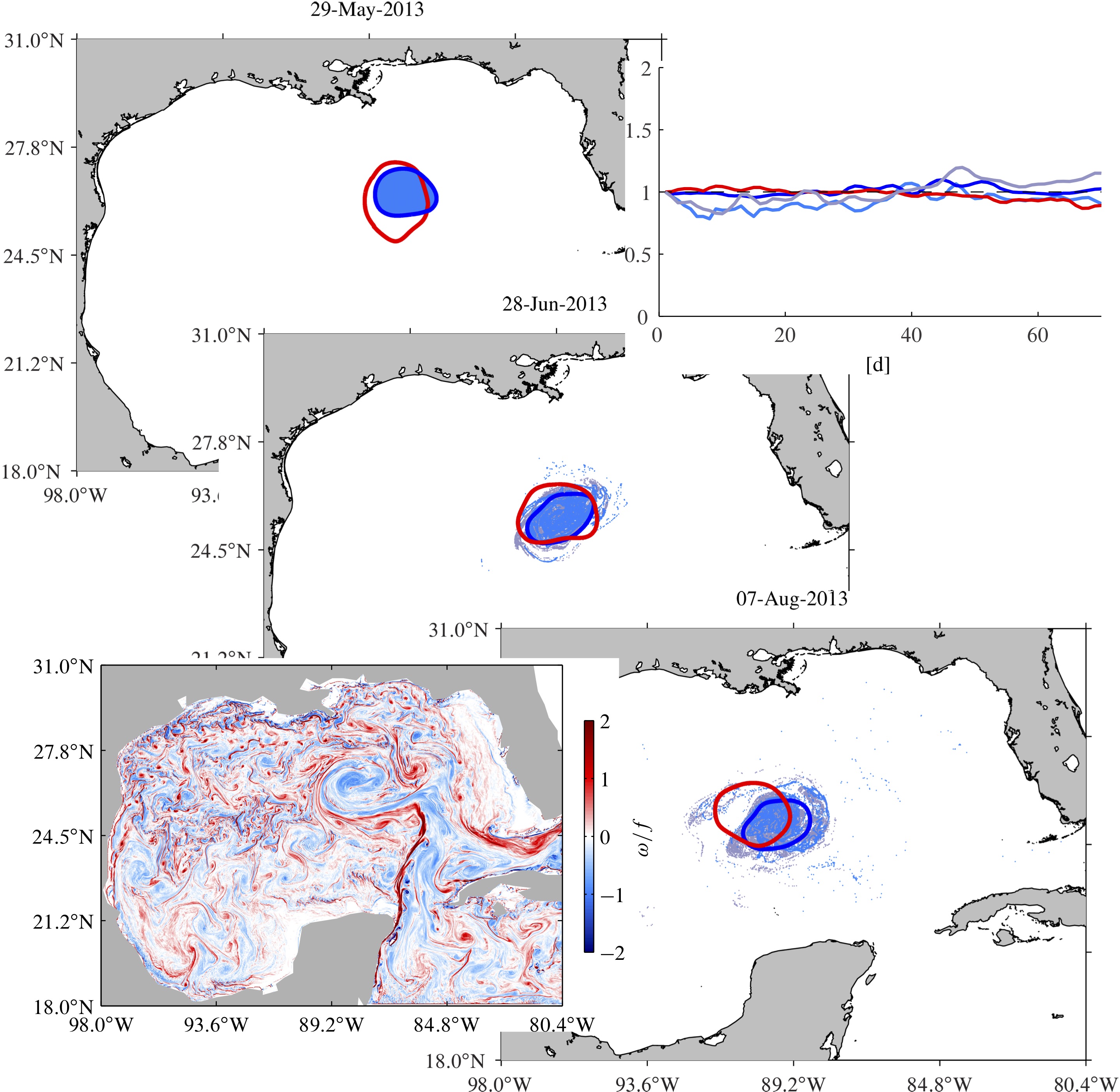}%
  \caption{Snapshots of the evolution of the boundary of coherent
  Larangian LCR \emph{Kraken} as extracted from altimetric SSH (red)
  and subsampled SSH from a submesoscale permitting Navy Coastal
  Model (NCOM) simulation of the GoM (blue), and a tracer patch
  passively advected by the total model velocity at the surface
  (light blue) and at 50-m depth (gray).  The top-right panel shows
  the mean distance to the corresponding centroid from points on
  the altimetry-derived boundary (red), the model-SSH-derived
  boundary (blue), and the boundary of the tracer patch passively
  advected by the total model velocity at the surface (light blue)
  and at 50-m depth (gray), each normalized by the value taken
  initially on 29 May 2013. The bottom-left panel is snapshot of
  the surface vorticity (normalized by the mean Coriolis parameter)
  field on 29 May 2013.   Figures constructed using MATLAB R2017b
  (http://www\allowbreak .mathworks.\allowbreak com/).}
  \label{fig:ncom}%
\end{figure}

It is still of interest to investigate if mesoscale coherent
Lagrangian eddies can be realized in the presence of submesoscale
motions.  We carried such an investigation by seeking evidence of
coherent Lagrangian LCR \emph{Kraken} in the data-assimilative
submesoscale-permitting NCOM simulation of the GoM.  The bottom-left
panel of Fig.\ \ref{fig:ncom} shows a snapshot of the surface
vorticity ($\omega$) normalized by the mean Coriolis parameter in
the GoM ($f$) from this simulation, which reveals a large number
of eddy-like features spanning scales in the meso- and submesoscale
ranges.  Note in particular the suprainertial ($|\omega|/f > 1$)
submesoscale eddies rolling around the mesoscale eddy feature in
the center of the GoM, which appears to be the signature of LCR
\emph{Kraken} in the model vorticity field.

However, we have not been able to extract a mesoscale coherent
Lagrangian ring akin to LCR \emph{Kraken} from the NCOM upper ocean
(surface, 10, and 50-m levels) velocity field on any day and for
any coherence timescale choice or stretching parameter.  But we
have found that a velocity field computed geostrophically using
subsampled model SSH does sustain one.  The procedure employed to
obtain the subsampled SSH field emulates that used to construct an
SSH field by interpolating along-track satellite altimetry measurements
\cite{Beron-JGR-10}.  Figure \ref{fig:ncom} shows snapshots of the
evolution of the extracted coherent Lagrangian boundary (dark blue)
along with the boundary obtained from satellite altimetry (red).
The area enclosed by the simulated ring boundary is 1.6 times
smaller than that enclosed by the observed boundary.  The centroid
of the simulated ring relative to that of the observed ring varies
from no less that 10 km to as much as 85 km over 80 d since the
extraction date. Over that period the observed ring translates
mainly westward, while the simulated ring translates mainly
southwestward.  The mean westward translation speed for the observed
ring is about 3.5 km~d$^{-1}$, while that of the simulated ring is
nearly 2 km~d$^{-1}$.  Note that shallow-water lenses on a $\beta$-plane
translate westward at a seeped of \cite{Nof-81a} $\beta L_\mathrm{D}^2
\approx 3.6$ km~d$^{-1}$ for a typical Rossby deformation scale in
the GoM of 45 km \cite{Chelton-etal-98}.  This agrees well with
the westward translation speed of the ring extracted from satellite
altimetry.  However, the ring extracted using simulated altimetry
translates westward more slowly, at nearly half the predicted speed.
Furthermore, the coherence horizon for the simulated ring is $T =
80$ d, which implies a much weaker long-range transport ability for
this ring than for the observed ring, whose coherence timescale is
$T = 200$ d.

\begin{figure}[p!]
  \centering%
  \includegraphics[width=\textwidth]{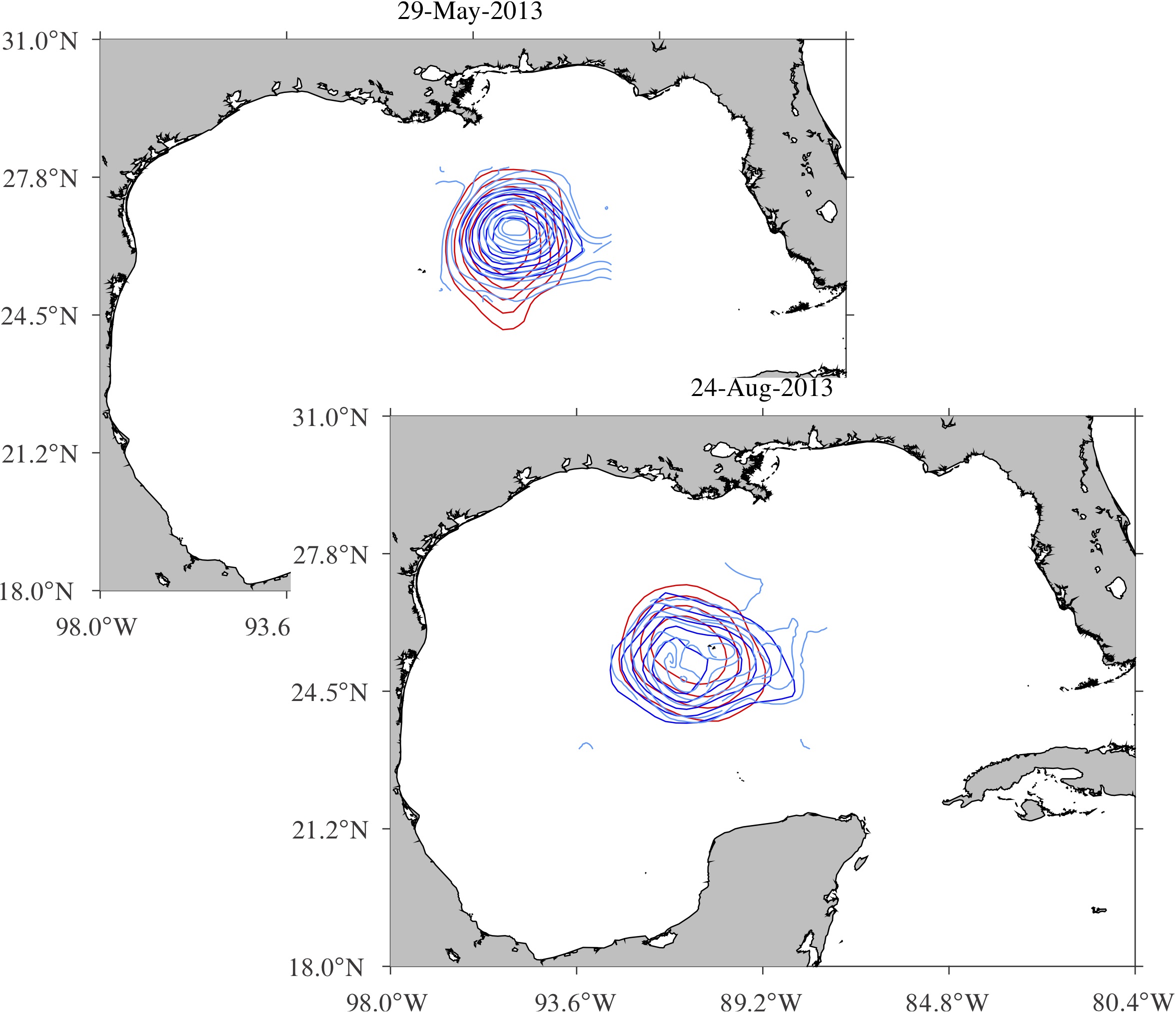}%
  \caption{Snapshots of altimetric-SSH-derived (red), model-SSH-derived
  (blue), and surface model velocity (light blue) streamlines.
  Figures constructed using MATLAB R2017b (http://www\allowbreak
  .mathworks.\allowbreak com/).}
  \label{fig:streamlines}%
\end{figure}

The above differences in Lagrangian behavior are not obvious from
naked-eye comparison of satellite and simulated altimetry SSH or
total model surface velocity streamlines.   Figure \ref{fig:streamlines}
shows snapshots of the latter in red, blue, and light blue,
respectively, on nearly three-month apart dates, restricted to the
vicinity of the region occupied by the observed LCR.  Satellite and
model SSH streamlines are closed, nearly concentric at all times.
In turn, total model surface velocity also show large (mesoscale)
nonintersecting streamline loops, which, unlike altimetry and model
SSH streamlines, can encircle smaller (submesoscale) closed as well
as open streamline loops.  A cursory Eulerian evaluation would
indicate mesoscale Lagrangian coherence in all cases simply because
the instantaneous streamlines remain closed as time progresses.  A
more elaborated Eulerian analyisis \cite{Flierl-81} would assess
the ability of these compact streamline flow regions to self advect
by measuring their degree of nonlinearity to rule out the possibility
that they represent purely linear wave superpositions.  A basic
estimate of the degree of nonlinearity of a wave is given by the
ratio $U/c$ of the maximal mean speed $U$ at the periphery of a
closed streamline region to its translation speed $c$.  When $U/c
> 1$, fluid may be expected to be carried along by a wave, and the
wave will in general be dynamically nonlinear, in that its
self-advection of momentum or vorticity affects its own evolution.
In all cases $U/c > 1$, inferring Lagrangian coherence over regions
of similar size.  But it is possible \cite{Haller-05} to construct
examples of flows with closed streamlines at all times for which
$U/c > 1$ in a given reference frame, while no coherent Lagrangian
eddy is present by observing the motion on a frame in which the
flow is steady.  Thus, given the observer-dependent nature of this
Eulerian analysis, it is not surprising that it leads to a coherence
assessment at odds with that drawn from the observer-independent
Lagrangian analysis carried out here.  The nonlinear dynamical
systems analysis provides unambiguous means for assessing the
performance of a model through various parameters that measure its
skill in representing observed Lagrangian behavior.  These metrics
include the various measures discussed above, namely, coherence
timescale, area enclosed, trajectory, and translation speed of
Lagrangian rings.

The differences in Lagrangian behavior found are puzzling given
that the NCOM simulation was shown to reproduce relative dispersion
statistics down into the marginally submesoscale range
\cite{Beron-LaCasce-16} as observed during the Grand LAgrangian
Experiment (GLAD) \cite{Olascoaga-etal-13}.  Irrespective of the
ability of the model to reproduce observed behavior, a valid question
still is whether the simulated total velocity field supports some
form of mesoscale Lagrangian coherence.  Recall that no coherent
Lagrangian ring was possible to be extracted from the total model
flow, neither at the surface nor at 10 m or 50 m.  The latter lies
below the model's mixed layer during the summer period analyzed,
which makes one wonder why velocities at that depth would not support
a coherent Lagrangian eddy boundary.  However, the mean kinetic
energy (about 0.065 m$^2$s$^{-2}$) is roughly uniform within the
upper 50 m, indicating similarly intense submesoscale activity at
the surface and below, inside this layer.  With this in mind, we
advected under the total model velocity, both at the surface and
50 m, passive tracers initially within the coherent Lagrangian
boundary detected from simulated altimetry.  This numerical experiment
revealed the existence of a swirling Lagrangian structure at the
surface (light blue dots in the diagonal panels of Fig.\ \ref{fig:ncom}),
which, while experiencing filamentation, prevailed in large part
confined within the interior of the boundary obtained from simulated
altimetry.  The total model flow at 50 m revealed a similar form
of coherence, as it can be seen in the same figure (gray dots).  A
compactness measure of these structures is provided in the top-right
panel of Fig.\ \ref{fig:ncom}.  The compactness measure considered
is given by the mean distance from the boundary of a shape with
respect to its centroid \cite{Haralick-74}.  The figure shows this
distance from points initially on 29 May 2013 lying along the
boundary extracted from satellite altimetry and as they are advected
by the flow derived from it (red), the boundary obtained from
simulated altimetry and as they are advected by the corresponding
velocity (blue), and the boundary of the tracer patch and as they
are advected by the total model velocity at the surface (light blue)
and at 50-m depth (gray), each normalized by the value taken on 29
May 2013.  Note that the compactness measure remains close to one
for the boundaries extracted from satellite and simulated altimetry,
consistent with their coherent Lagrangian nature.  The departure
from one can be larger for the tracer clouds advected by the total
model velocity at the surface and 50 m, but this is never in excess
of 0.25.

Important steps toward rigorously characterizing the reported form
of mesoscale Lagrangian coherence amid submesoscale motions have
been made in recent work by Beron-Vera et al.\ \cite{Beron-etal-18a}.
Lagrangian coherence in that work is assessed based on the ability
of the elements of a material loop to rotate in consonance
\cite{Haller-etal-16} or a region to maintain short distances among
themselves relative to their distances to particles outside of the
region \cite{Hadjighasem-etal-16}.  Clearly, the relevance of such
a form of Lagrangian coherence depends on the extent that the
intensity of the submesoscale activity in the model can be supported
with observations.  The observations analyzed here do not provide
support for it in the specific region of the GoM and timespan
considered.

\section*{Summary and discussion}

In this paper we have provided independent observational support
for the enduring Lagrangian coherence of a Loop Current ring (LCR)
in the Gulf of Mexico (GoM), which was detected from its Eulerian
footprints in satellite altimetry sea surface height (SSH) data and
termed \emph{Kraken}.  Coherent Lagrangian eddies have material
boundaries that do not filament or experience global break away
over the coherence assessment time interval.  A recent technique
rooted in nonlinear dynamical systems theory, geodesic eddy detection,
has been specifically designed to unveil these eddies from a velocity
realization in an observer-independent fashion.

Applying geodesic eddy detection on the altimetry-derived velocity
field, we inferred that LCR \emph{Kraken} possessed a Lagrangian
core of about 100-km radius, surrounded by a resilient boundary
that remained so for a period as long as seven months.  Lagrangian
boundaries with similar but shorter-lived resilience, which enclosed
a larger domain including this core, were found to emerge over this
period.  These boundaries conveyed further Lagrangian coherence to
LCR \emph{Kraken}.

The observational support for the altimetry-inferred material
coherence was obtained from two, rarely available simultaneously,
independent sources: satellite ocean color and satellited-tracked
surface drifter trajectories. A low-concentration chlorophyll patch
traceable into the Caribbean Sea accompanied LCR \emph{Kraken} in
its nearly steady westward translation across the GoM.  The surface
drifters, in turn, developed relatively wiggle-free anticyclonic
loops inside or immediately outside LCR \emph{Kraken}'s Lagrangian
core with turnover times (of two weeks or so) very close to those
of passive tracer particles circling along the boundary of the core.

Recent eddy retention studies \cite{dOvidio-etal-13, Condie-Condie-16,
Lehahn-etal-11, Gaube-etal-14, Moreau-etal-17} suggest that the
reported Lagrangian supercoherence for LCR \emph{Kraken} may be
extended to rings and eddies in other regions of the world oceans.
This clearly requires an investigation that exceeds the breadth of
the present contribution.

The output from the operational U.S. Navy Coastal Ocean Model (NCOM)
of the GoM was finally analyzed in search for evidence of the
observed coherent Lagrangian LCR.  Neither the flow derived using
subsampled model SSH simulating satellite altimetry nor the total
upper-ocean model flow was found to support a coherent Lagrangian
ring with the same characteristics as the observed coherent Lagrangian
LCR. The simulated altimetry revealed a coherent Lagrangian ring
which was smaller and much shorter lived than the observed ring,
and the total model flow did not support a coherent Lagrangian ring
at all, neither at the surface nor within the first 50~m below of
it.  However, swirling Lagrangian structures, which can be characterized
as coherent but in a relaxed sense \cite{Beron-etal-18a}, were found
to be sustained by the total upper-ocean model flow.

The lack of agreement between the simulation and the observed
behavior as evidenced by the low model skill as measured by the
several Lagrangian metrics resulting from the application of the
nonlinear dynamical systems tools is puzzling.  On one hand, an
altimetry-assimilative system that marginally resolves submesoscales
should be able to capture observed mesoscale motions with a reasonable
level of accuracy.  On the other hand, the model was recently found
\cite{Beron-LaCasce-16} to reproduce observed relative dispersion
statistics correctly down into to the marginally submesoscale range.

A possible explanation for the undesired behavior of the NCOM
simulation may lie in the way that altimetry data are assimilated
into the model, which was found \cite{Jacobs-etal-14} to produce
poor agreement between observed and simulated instantaneous Lagrangian
transport patterns.  The agreement was improved \cite{Jacobs-etal-14}
by dividing the analysis increment fields by the data time window
length to more properly weight the data.  However, to the best of
our knowledge,  this fix has not yet been implemented operatively,
and is mainly expected \cite{Jacobs-etal-14} to lead to stronger
current fields, which is unclear how they can impact tracer dispersion
over time.  Another explanation may be found in the increased
resolution of the model toward the submesoscale, which can conspire
against its mesoscale forecasting skill as it has been recently
shown \cite{Sandery-Sakov-17}.  Discerning among these explanations
and possibly additional causes clearly deserves investigation, but
this is beyond the scope of this paper.  
 

\section*{Acknowledgments}

The altimeter products were produced by SSALTO/DUCAS and distributed
by AVISO with support from CNES (http://\allowbreak www.\allowbreak
aviso.\allowbreak oceanobs).  The ocean color products were produced
and distributed by NASA OBPG.  The surface drifters were deployed
as part of Horizon Marine Inc.'s EddyWatch\textsuperscript{\textregistered}
program.  The data have been provided as a part of a data exchange
agreement between Horizon Marine Inc.\ and PEMEX under CICESE--PEMEX
SAP contracts 428217896, 428218855, and 428229851. The quality
control and post-processing of these data were carried out by Paula
Garc\'{\i}a and Argelia Ronquillo. The NCOM simulation was produced
at the Naval Research Laboratory and can be obtained from the Gulf
of Mexico Research Initiative Information \& Data Cooperative
(GRIIDC) at https://data.\allowbreak gulfresearchinitiative.\allowbreak
org (dx.doi.org/\allowbreak 10.7266/\allowbreak N7FQ9TJ6,
dx.doi.org/\allowbreak 10.7266/\allowbreak N76Q1V5G, and
dx.doi.org/\allowbreak 10.7266/\allowbreak N72Z13F4).  Support for
this work was provided by the Gulf of Mexico Research Initiative
(MJO, FJBV, and YW) as part of the Consortium for Advanced Research
on Transport of Hydrocarbon in the Environment (CARTHE); SENER--CONACyT
grant 201441 (MJO, FJBV, PPB) as part of the Consorcio de Investigaci\'on
del Golfo de M\'exico (CIGoM); and NOAA/AOML and UM/CIMAS (JT).
`
\section*{Author Contributions}

MJO, FJBV and YW performed the dynamical systems analysis, JT the
ocean color data analysis, and PPB the drifter trajectory data
analysis.  All authors contributed to the interpretation of the
results and the writing of the manuscript.

\section*{Additional Information}

\noindent\textbf{Competing interests:} The authors declare
no competing interests.

\newpage

\end{document}